\documentclass[prl,twocolumn,showpacs,superscriptaddress,preprintnumbers,amssymb]{revtex4}
\usepackage{graphicx}
\usepackage{dcolumn}
\usepackage{bm}
\usepackage{wasysym}
\usepackage{psfrag}
\usepackage{subfigure}
\usepackage{sidecap}
\newcommand{\figurewidth}{84mm}

\begin{document}
\title{ Origin of large moments in Mn$_x$Si$_{1-x}$ at small x}
\author{M. Shaughnessy}
\affiliation{Department of Physics, University of California, Davis, CA 95616-8677}
\author{C.Y. Fong}
\affiliation{Department of Physics, University of California, Davis, CA 95616-8677}
\author{Ryan Snow}
\affiliation{Department of Physics, University of California, Davis, CA 95616-8677}
\author{Kai Liu}
\affiliation{Department of Physics, University of California, Davis, CA 95616-8677}
\author{J.E. Pask}
\affiliation{Condensed Matter and Materials Division, Lawrence Livermore National Laboratory, Livermore, CA 94551}
\author{L.H. Yang}
\affiliation{Condensed Matter and Materials Division, Lawrence Livermore National Laboratory, Livermore, CA 94551}
\date{\today}
\begin{abstract}
Recently, the magnetic moment/Mn, $M$, in Mn$_x$Si$_{1-x}$ was measured to be 5.0 $\mu_B$/Mn, at $x$ =0.1\%. To understand this observed $M$, we investigate several Mn$_x$Si$_{1-x}$ models of alloys using first-principles density functional methods. The only model giving $M = 5.0$ was a 513-atom cell having the Mn at a substitutional site, and Si at a second-neighbor interstitial site. The observed large moment is a consequence of the weakened d-p hybridization between the Mn and one of its nearest neighbors Si, resulting from the introduction of the second-neighbor interstitial Si. Our result suggests a new way to tune the magnetic moments of transition metal doped semiconductors.


\end{abstract}


\maketitle

Spintronic devices, which exploit the electron spin as well as its charge, hold the promise of breakthroughs for sensors, memory chips, microprocessors, and a host of other applications \cite{Ohno,Ball}. Among the candidate materials for spintronic devices, dilutely doped transition elements in III-V compounds have been among the most studied. However, issues of sample quality \cite{Reed} and low Curie temperature \cite{Ball} have limited their usefulness for spintronic applications in practice. Recently, dilute alloys of Mn$_x$Si$_{1-x}$ have attracted attention after reports \cite{Nakayama, Park} of epitaxially grown thin films of Mn$_x$Si$_{1-x}$ with $x = 5.0$\% that exhibit the anomalous Hall effect at $\sim$70~K. The demonstration of magnetic properties in Si-based materials and the greater maturity of Si-based technologies relative to those of other semiconductors enhance the prospects for utilizing Mn$_x$Si$_{1-x}$ to realize spintronic devices. Among more recent experiments on Si-based alloys \cite{Zhang,Bolduc,Ma}, Refs.~\cite{Bolduc} and \cite{Ma} report the magnetic moment/Mn, $M$. At $x =0.1$\%, an $M$ of 5.0 $\mu_B$/Mn is found. This value is the maximum achievable when all five d-electron spins align and therefore is particularly interesting that it occurs in the solid alloy. A number of calculations \cite{Dalpian,da Silva,Bernardini,Weng, ZZhang} have been carried out with Mn in a variety of configurations in various supercells. The total number of atoms in the supercells range from 32 to 218. Only when imposing a 2+ charge state on the Mn has the observed $M = 5.0 \mu_B$/Mn been reproduced \cite{Bernardini}. However, during the ion implantation \cite{Bolduc} and arc melting \cite{Ma} growth of dilute Si-based alloys, no particular charge state of the Mn is prepared. Therefore, we consider the following questions: Is the Mn$^{2+}$ charge state necessary to obtain the maximum $M$ in Si based alloys? If not, how can such a large $M$  be obtained in an alloy? In this Letter, we show for the first time that the large, experimentally observed 5.0 $\mu_B$/Mn moment in Mn$_x$Si$_{1-x}$ can in fact be achieved, without charge-state or other such constraints, in the presence of interstitial defects at sufficiently low Mn concentrations.

\begin{figure}
  \includegraphics[angle=0,width=2in,clip=true]{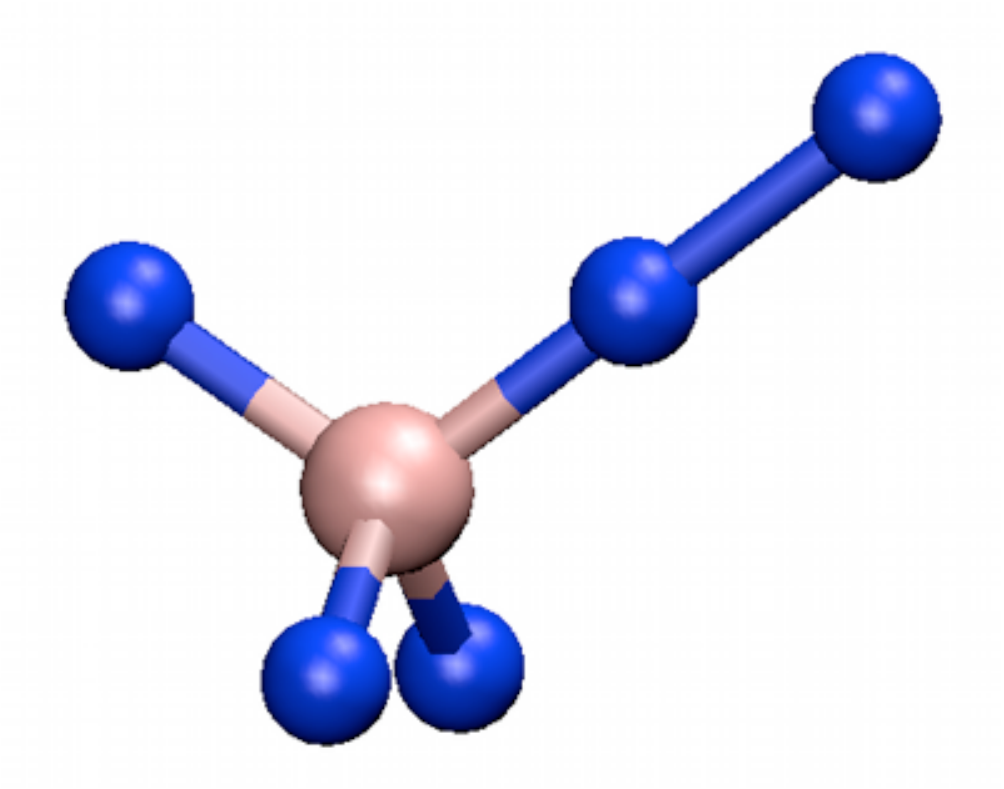}
  \caption{The Mn and its neighboring Si atoms. The Mn atom is the large (lighter) central sphere and the Si atoms are smaller (darker) spheres. The sn Si of the Mn is shown in the upper right at the TI site, and the ``nn Si'' is shown between it and the Mn. The other atoms in the 513-atom cell are omitted for clarity.}
 \label{fig:512_snapshot}
\end{figure}

 We carried out extensive {\it ab initio} density functional calculations on a range of supercell models. Each model has a single Mn atom doped into a 64-, 216- or 512-Si cell. The Mn occupies a substitutional (S), tetrahedral interstitial (TI) or hexagonal interstitial (HI) site. Among all the models, only one was found to have the observed 5.0 $\mu_B$/Mn: a 513-atom cell with substitutional Mn and interstitial Si at a second-neighbor (sn) TI site. 
 The optimized lattice constant is 5.46\AA\space, compared to the experimental value of 5.43\AA. A planewave basis and ultrasoft pseudopotentials \cite{VASP} were employed, with Perdew91 GGA exchange-correlation \cite{Perdew}, 650 eV planewave cutoff, and (3,3,3) Monkhorst-Pack $k$-point mesh. In Fig.~\ref{fig:512_snapshot}, we show the portion of the supercell containing the Mn, its nearest neighbor (nn) Si's, and second-neighbor (sn) Si interstitial after all atoms are relaxed to forces less than 6.0 meV/\AA. Hereafter, we refer to the nn Si between the Mn and sn Si as the ``nn Si" due to its unique role in the present context.
 


The unique features of this 512-atom model are: (1) It corresponds to x = 0.19\%, within a factor of 2 of the concentration in the experiments \cite{Bolduc}, (2) the impurity atom and its sn Si are well isolated from neighboring periodic images, and (3) it gives the magnetic moment 5.0 $\mu_B$/Mn. Feature two is demonstrated by the fact that before relaxation, the maximum component of the force at the most distant Si from the Mn is already small at 0.015 eV/\AA. In the corresponding 217-atom cell, the corresponding components are 0.08 eV/\AA.

According to the ionic model \cite{Fong}, a Mn atom at an S site in Si should have a magnetic moment of 3.0 $\mu_B$/Mn owing to four of its seven electrons hybridizing with the four neighboring Si atoms. The three remaining electrons at the Mn align their spins in accordance with Hund's first rule to give the stated magnetic moment. In the 512-atom cell with no sn Si (nsn), the calculated moment is in fact exactly 3.0 $\mu_B$/Mn. 

\begin{figure}[htp]
\includegraphics[angle=0,width=\figurewidth,clip=true]{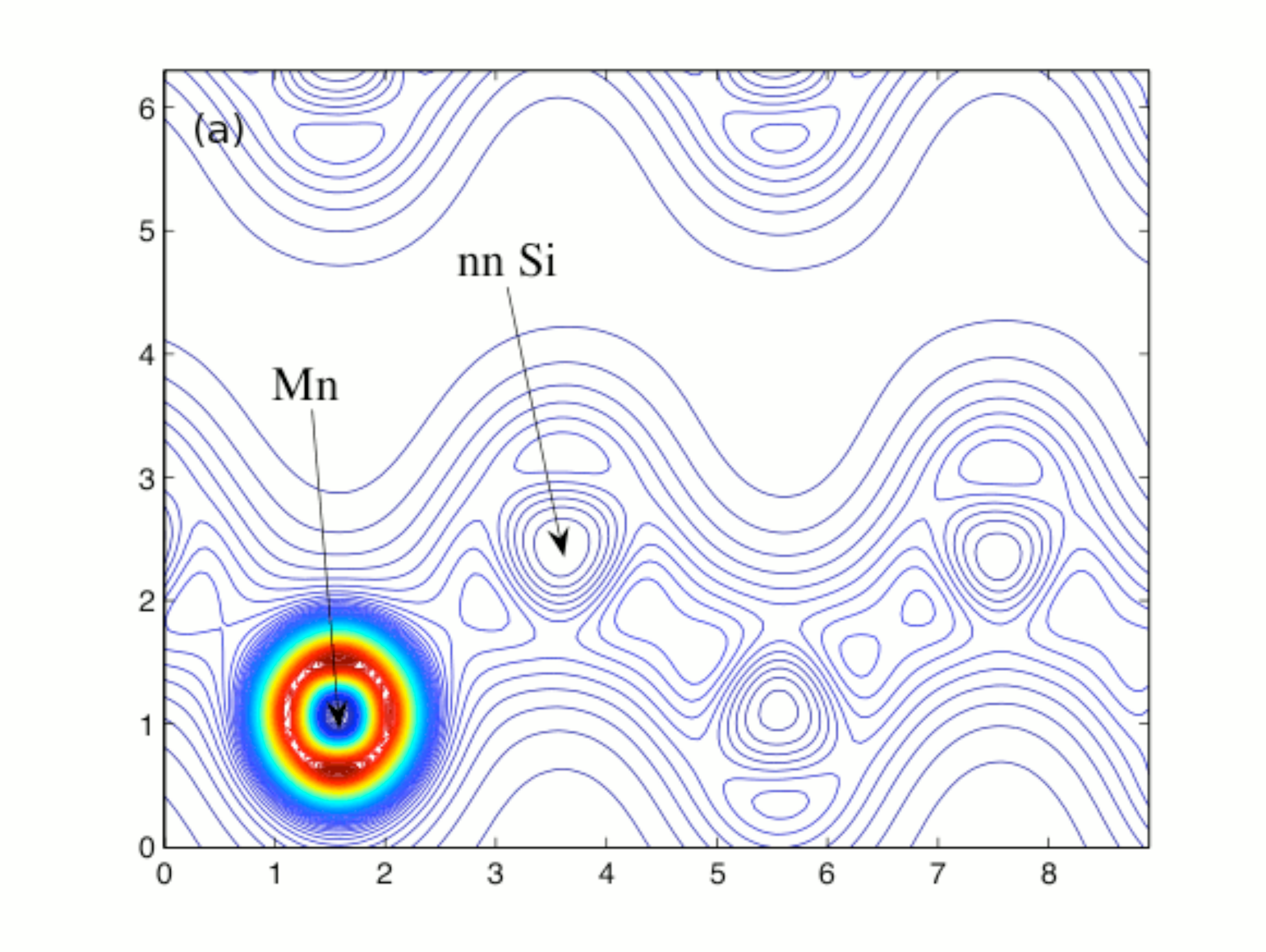}

%
  \includegraphics[angle=0,width=\figurewidth,clip=true]{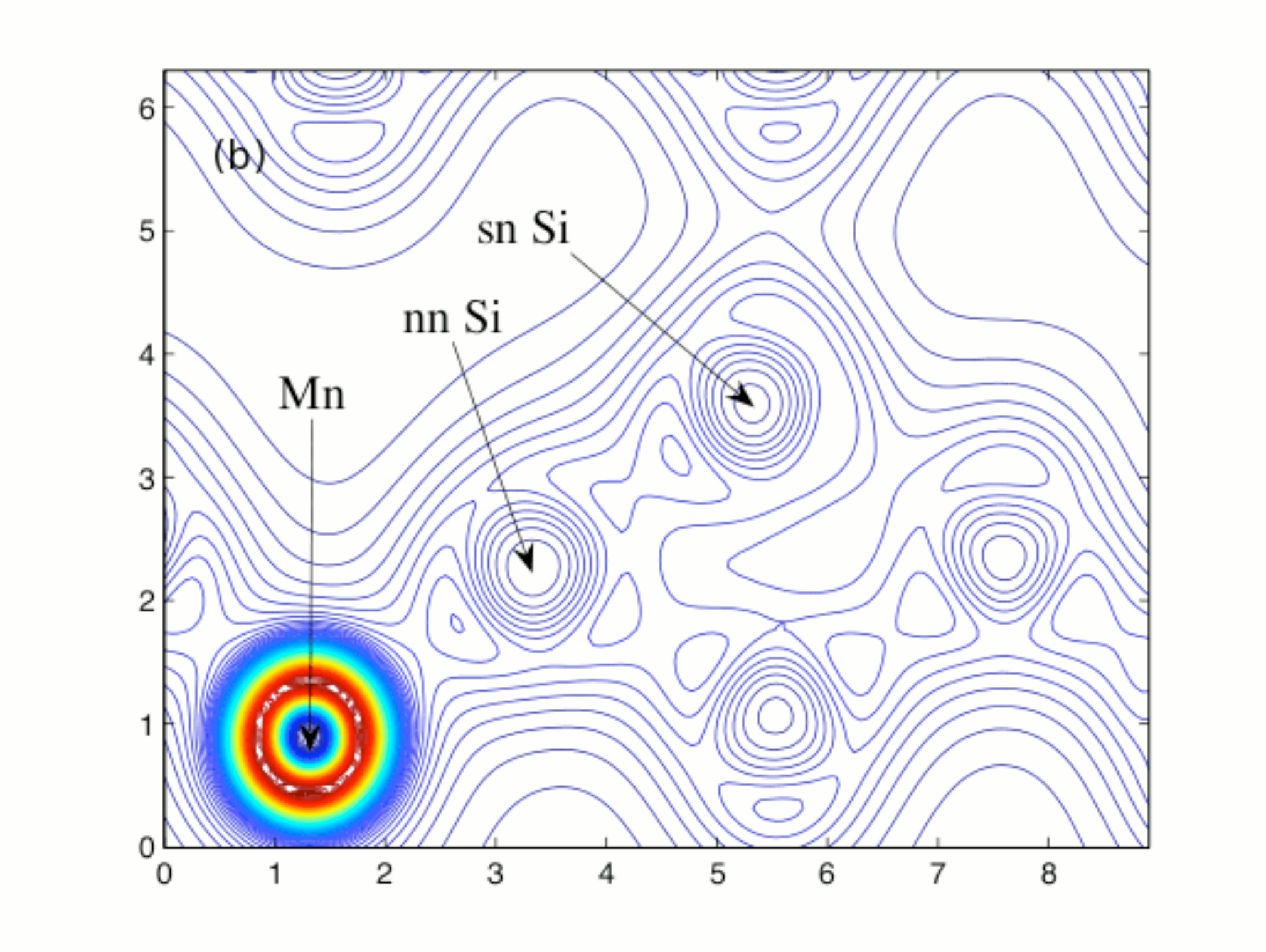}
  \includegraphics[angle=0,width=\figurewidth,clip=true]{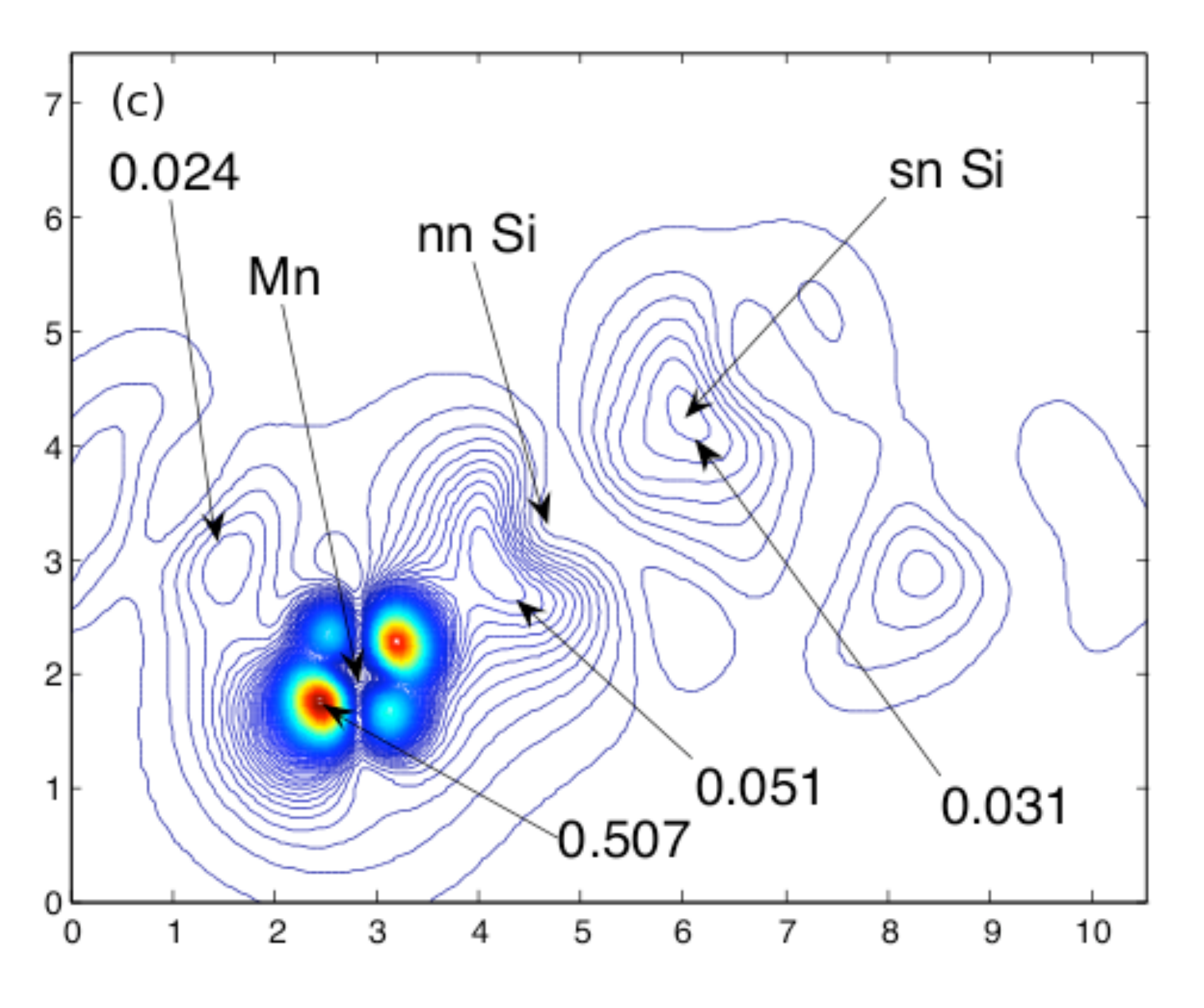}

           
 \caption{The charge density in the majority (down) spin channel for (a) the nsn model and (b) the sn model. The bottom panel (c) shows the spin density difference as given in Eq (1), with the centers of the Mn aligned. The labels show the positions of the relevant atoms and selected contour values ($e$/\AA$^3$). All distances are in \AA. The $\bf{x}$-axis of the partial cross-section is the [110] direction of the supercell and the $\bf{y}$-axis coincides with the $\bf{z}$-axis of the supercell.}  
\end{figure}

The bonding between the nn Si and its three Si neighbors is of sp$^3$ type while that between the nn Si and Mn shows strong d-p hybridization. Figure 2(a) shows the charge density of the majority (down) spin channel in the vicinity of the Mn and nn Si in the nsn model. The minority (up) spin charge density (not shown) is qualitatively similar. The dense (brightly colored online) contours in the bottom left of the figure correspond to the large d-electron density in the vicinity of the Mn. The d-p hybrid bonding between the Mn and nn Si differs from the bonding between the nn Si and its other neighbors, as manifested in the distinct charge distributions between the atoms. The covalent bonds between Si atoms exhibit the characteristic dumbbell shape, with maxima near host atoms.  
Figure 2(b) shows the charge density in the same spin channel as in 2(a) but for the sn model. In the presence of the sn Si, the sp$^3$ orbital of the nn Si that participated in the d-p bonding is {\it pulled away} from the Mn, thus reducing the hybridization. The charge distribution between the nn Si and sn Si shows an asymmetric dumbbell shape indicating that the nn Si shifts some of its charge toward the sn Si and away from the Mn. When the bond between the Mn and nn Si is weakened in the presence of the sn Si, the two electrons in the bond retreat toward their respective host atoms. The maximum charge densities along the bond between the Mn and nn Si 
in front of and behind the Mn atom, increase by $\sim$13\% and $\sim$9\%, respectively, relative to the case without the sn Si. Along the same line, the maximum nearer to the nn Si increases by $\sim$5\% and moves closer to the nn Si by $\sim$7\%, relative to the nsn model. From an atomic point of view, the addition of the sn Si reduces the hybridization between the Mn d-orbital and the nn Si sp$^3$ states. The electron retreating to the Mn aligns its spin in accordance with Hund's rule and a ferromagnetic exchange polarizes the density around the nn Si. 


To quantify the weakening of the d-p hybridization, we computed the density of d-states (d-DOS) in the down spin channel with and without the sn Si and computed the variances of the distributions. The variance of d-DOS for the sn model was found to be 4.08 eV$^2$ while that for the nsn model was 4.53 eV$^2$ in an energy window from E$_f$ to 3.0 eV below. The narrower width of the d-DOS in the presence of the sn Si confirms the weakening of the d-p hybridization suggested by the corresponding charge distribution.


The difference of total spin up and spin down charge is $3e$ and $5e$ in the nsn and sn cases, respectively. 
As a consequence of the weakened d-p hybridization, the local moment around the Mn is about 4 $\mu_B$. The residual weak spin polarization in the rest of the unit cell contributes approximately 1 $\mu_B$ to the total moment and consequently, the $M$ value for the relaxed dilute Mn$_x$Si$_{1-x}$ is 5.0 $\mu_B$/Mn in agreement with the measured value after the sample is annealed \cite{Bolduc}.


Figure 2(c) illustrates the effect of the sn Si on the magnetic moment distribution by showing the difference $\Delta \sigma({\bf r})$ of the spin densities with and without the sn Si. The spin density is defined as the difference between the up and down charge densities: $\sigma({\bf r}) = \rho({\bf r})_{down} - \rho({\bf r})_{up}$
The contours are determined by the difference of spin densities
\begin{equation}
\Delta \sigma({\bf r})= \sigma({\bf r})_{sn} - \sigma({\bf r})_{nsn},
\end{equation}
where first and second terms on the right hand side are related to the densities with and without the sn Si, respectively. The spin densities derived from Figs.~2(a) and 2(b) are not directly subtracted because the atomic positions differ upon relaxation. Instead, both densities are determined with the sn model atomic positions. 
The distribution near the Mn shows the four lobes of a d-state with some of its charge shifted from the lobe pointing to the nn Si back to the open region behind the fourfold-coordinated Mn atom. The dense region (brightly colored online) in the lower left, shows the strong positive $\Delta \sigma({\bf r})$ due to the  d-electron's retreat as the d-p hybridization is weakened. The (blue online) side contours near the nn Si indicate an outwards displacement of the sp$^3$ orbital into the open space around the bond after it retreats to the nn Si. 
The overlap of the wavefunctions as the sn Si shifts some of its charge toward the nn Si induces some moment near the sn Si, indicated by the contours in Fig.~2(c) near the sn Si. The labeled values of the contours in Fig.~2(c) show the dominant effect is a large change in the immediate vicinity of the Mn atom with smaller changes in moment near the nn and sn Si. The reduced d-p hybridization allows for the sp$^3$ electron of the nn Si to spill out into the empty region surrounding the bond, reducing the spatial overlap of the d and sp$^3$ electrons and making their spin alignment more energetically favorable.     

In summary, we found the experimentally observed 5.0 $\mu_B$/Mn moment in Mn$_x$Si$_{1-x}$ with $x =0.1$\% can only be accounted for, among all models considered, by an isolated Mn at an S site coupled with sn Si at a TI site in a sufficiently large unit cell. The concentration $x$ of the unit cell considered corresponds to 0.19\%, about a factor of two larger than experiment. Based on the mechanism proposed it is predicted that smaller concentrations, corresponding to more computationally demanding models, will yield precisely the same $M$. 
The picture which emerges for the 5.0 $\mu_B$/Mn moment in Mn$_x$Si$_{1-x}$ is one in which the sp$^3$ orbital of the nn Si participating in the bonding with the Mn is pulled away from the Mn, reducing the d-p hybridization and leaving the Mn in a higher-spin, more atomic-like state. The polariztions about the Mn, nn Si, and sn Si all contribute to the measured moment, with the Mn atom giving the dominant contribution. It will be of interest to consider this mechanism of moment enhancement in other dilute magnetic semiconducting and/or half metallic systems. 

This work was supported by the NSF Grant No. ECCS-0725902. Work at Lawrence Livermore National Laboratory was performed under the auspices of the U.S. Department of Energy under Contract DE-AC52-07NA27344.


\end{document}